\documentclass[twocolumn,amsmath,amssymb,aps,prc]{revtex4-2}

\usepackage{graphicx,here}
\usepackage{dcolumn,bm,hyperref,footnote}

\usepackage{float}
\usepackage[dvipsnames,svgnames,x11names]{xcolor}
\usepackage[authormarkup=none,defaultcolor=BrickRed]{changes}

\begin{document}

\title{Stability of the manifold boundary approximation 
method for reductions of nuclear structure models}

\author{M. Imbri{\v s}ak}
\affiliation{Department of Physics, Faculty of Science, 
University of Zagreb, HR-10000 Zagreb, Croatia}

\author{K. Nomura}
\email{knomura@phy.hr}
\affiliation{Department of Physics, Faculty of Science, 
University of Zagreb, HR-10000 Zagreb, Croatia}

\date{\today}

\begin{abstract}
The framework of nuclear energy density 
functionals has been employed 
to describe nuclear structure phenomena 
for a wide range of nuclei. 
Recently, statistical properties 
of a given nuclear model, such as parameter 
confidence intervals and correlations, 
have received much attention, particularly 
when one tries to fit complex models. 
We apply information-theoretic methods 
to investigate stability of model reductions 
by the manifold boundary 
approximation method (MBAM). 
In an illustrative example of 
the density-dependent point-coupling model 
of the relativistic energy density functional, 
utilizing Monte Carlo simulations, 
it is found that main conclusions obtained 
from the MBAM procedure are stable
under variation of the model parameters.
Furthermore, we find that the end of the 
geodesic occurs when the determinant 
of the Fisher information metric vanishes, 
thus effectively separating the parameter 
space into two disconnected regions. 
\end{abstract}

\maketitle

\section{Introduction}

The nuclear energy density functional (EDF) framework  
is a promising, unified theoretical approach for a global 
description of nuclear structure phenomena. 
One of the successful EDFs has been the one 
that is based on the relativistic mean-field Lagrangian 
in the finite-range meson-exchange model 
\citep{Bender2003Self-consistentStructure}, 
with the density-dependent meson-nucleon couplings 
providing an improved description of asymmetric 
nuclear matter \citep{VRETENAR2005RelativisticStructure}. 
Moreover, it has been found that simpler, 
point-coupling models 
\citep{Nikolaus1992NuclearModel,Burvenich2002NuclearModel} 
produce comparable results to the finite-range ones, 
even if the point-coupling interactions are being 
adjusted to nuclear matter and ground-state properties 
of finite nuclei \citep{Niksic2008Finite-Interactions}. 
These density-dependent point-coupling models, 
however, have been shown to exhibit an exponential 
range of sensitivity to parameter variations, 
prompting the application of model reduction methods based 
on concepts of information geometry 
\citep{Niksic.2016,Niksic.2017}.

The universal nuclear energy density functional 
project was a large-scale collaborative 
effort primarily focused on a wide range of pioneering 
developments in EDF, including 
uncertainty quantification of nuclear theory 
\citep{Kor.10,Bogner2013ComputationalProject}. 
In the last decade, statistical error analysis, 
employing either classical or Bayesian inference, 
has started to be recognized in EDF research for 
its ability to quantify theoretical errors, distinguish 
safe  and risky extrapolations, provide sensitivity 
analysis, and offer insight into model instabilities 
\citep{Reinhard2015EstimatingModels,Dobaczewski2014ErrorGuide,Schunck2014ErrorTheory,Schunck2015UncertaintyTheory,Piekarewicz2014InformationPhysics, Roca-Maza2015CovarianceInstabilities,Agbemava2019PropagationProperties,Dedes2018PredictiveCapacities,Dedes2019PropagationNuclei,Kejzlar2020StatisticalModels,Taninah2020ParametricFunctionals}.

Information geometry is an interdisciplinary field 
that introduces differential geometry concepts 
to statistical problems 
\citep{Amari1982,Amari2016InformationApplications} 
with its initial applications centered around 
machine learning and neural networks 
\citep{Amari1998NaturalLearning, Ma1997AnNetworks}. 
Recently, the manifold boundary approximation method (MBAM) 
\citep{Transtrum.2010,Transtrum.2011,Transtrum.2014} 
has been developed to study complex and sloppy problems 
occurring in physics, chemistry, and biology 
\citep{Transtrum.2015,Transtrum.2016,Tisanic2020TheNuclei} 
in order to either classify or reduce complex models, 
such as the nuclear EDFs 
\citep{Niksic.2016,Niksic.2017,FrancisUnwindingSystems}.

The complexity of nucleon-nucleon interaction 
in the nuclear medium, coupling between single-nucleon
and collective degrees of freedom, 
and finite-size effects present obstacles 
to numerous attempts to establish a 
single theoretical framework to treat the nuclear 
many-body problem. The nuclear EDFs, 
and structure models based on them, 
have become a promising tool for the 
description of ground-state properties and 
low-energy collective excitation 
spectra of medium-heavy and heavy nuclei. 
A variety of structure phenomena have 
been successfully described 
using the nuclear EDF framework with a high 
level of global precision and accuracy over the 
entire chart of nuclides, and at a very moderate 
computational cost.

The unknown exact nuclear EDF is approximated 
by functionals of powers and gradients of 
ground-state nucleon densities and currents, 
representing distributions of matter, spin, isospin, 
momentum, and kinetic energy. 
A generic density functional is not necessarily microscopic; 
i.e., it is related to the underlying internucleon 
interactions, but some of the most successful 
functionals are entirely empirical.
However, one can also follow the middle way between 
fully microscopic and entirely empirical EDFs, and
consider semiempirical functionals that start from 
a microscopically motivated ansatz for the nucleonic 
density dependence of the energy of a system of protons 
and neutrons. 
Most of the parameters of such a functional are adjusted, 
in a local density approximation, to reproduce a given 
microscopic equation of state (EoS) of 
infinite symmetric and asymmetric nuclear matter, 
and eventually neutron matter. 
The remaining, usually few, terms that do not 
contribute to the energy density at the nuclear matter level 
are then adjusted to selected ground-state data 
of an arbitrarily large set of spherical and/or deformed nuclei. 
A number of semiempirical 
functionals have been developed over the last decade
\cite{TW.99,Fin.04,Fin.06,Baldo.08,Baldo.13,NVR.08,NVR.11,Kor.10,Kor.12,Kor.14,Bul.15} and have been very successfully applied to 
studies of a diversity of structure properties, 
from clustering in relatively light nuclei to the 
stability of superheavy systems, and 
from bulk and spectroscopic properties of stable nuclei 
to the physics of exotic nuclei at the particle drip lines.

In the previous studies~\cite{Niksic.2016,Niksic.2017}, 
concepts from information geometry have been used 
to demonstrate that nuclear EDFs are, in general, ``sloppy'' 
\cite{Transtrum.2015,Transtrum.2014,Gutenkunst.2007,Transtrum.2010,Transtrum.2011}.
The term ``sloppy'' refers to the fact that 
the predictions of nuclear EDFs and related models 
are really sensitive to only a few 
combinations of parameters ({\it stiff} parameter combinations) 
and exhibit an exponential decrease of sensitivity 
to variations of the remaining combinations of 
parameters ({\it soft} parameter combinations). 
This means that the soft combinations of
parameters are only loosely constrained by the available 
data, and that most nuclear EDFs in fact contain models 
of lower effective dimensionality associated with 
the stiff combinations of model parameters. 
In Ref.~\cite{Niksic.2016}, 
the most effective functional
form of the density-dependent coupling parameters 
of a representative model EDF 
have been deduced by employing the MBAM \cite{Transtrum.2014}. 
The data used in this calculation
included a set of points on a microscopic EoS 
of symmetric nuclear matter and neutron matter. 
This choice was motivated by the necessity 
to calculate the derivatives of observables with respect 
to model parameters which is, of course, more 
easily accomplished for nuclear matter in comparison 
to finite nuclei. 
In Ref.~\cite{Niksic.2017},  
this calculation has been extended, by employing a simple numerical 
approximation, to calculate the derivatives of observables 
with respect to model parameters, thus allowing one 
to apply the MBAM to realistic models 
constrained not only by the nuclear matter EoS but
also by observables measured in finite nuclei.
During the analysis of parametrizations 
in Ref.~\cite{Niksic.2017} it has been found that the 
numerical integration of the geodesic equation 
could reach the manifold boundary in a finite number 
of integration steps, indicating the divergence of the metric 
tensor determinant in a particular region of 
the parameter space. This surprising behavior 
has motivated us to investigate the stability 
of model reductions obtained by the MBAM, 
since the divergent region might be unintentionally 
missed by using too large integration steps.

In this work, we study the stability of the MBAM 
with respect to the variation of the model parameters. 
In Sec.~\ref{sec:info-geom}, 
we give an introduction to information-geometric concepts. 
In Sec.~\ref{sec:numerics} 
we describe the numerical implementation 
for finding the Dirac mass and binding energies, 
aided by algorithmic differentiation. 
The results of our investigation are given 
in Sec.~\ref{sec:stability}, 
while further applications of information 
geometry to nuclear EDFs are discussed 
in Sec.~\ref{sec:conclusion}.

\section{Information geometry and model reduction\label{sec:info-geom}}

A selection of the model is usually made 
through the maximum likelihood method, 
with the assumption that at the $a$th measurement 
the data $(x^a,y^a)$ can be described using a normal 
distribution, denoted by $\mathcal{N}$, 
by a model function 
$f(x^a,\mathbf{p}) \equiv f^a(\mathbf{p})$ as 
$y^a\sim \mathcal{N}\left(f^a(\mathbf{p}),\left(\sigma^a\right)^2\right)$. 
Here, $\sigma^a$ is the uncertainty of 
each measurement, and $\mathbf{p}$ is chosen 
from an appropriate parameter space, 
denoted by $\mathcal{M}$. 
Finding the best-fitting model is equivalent 
to maximizing the following log-likelihood function 
$l(\mathbf{p})$ over $\mathbf{p}\in\mathcal{M}$, 
\begin{equation}
 l(\mathbf{p})=\sum\limits_a \ln\phi\left(\frac{y^a-f^a(\mathbf{p})}{\sigma^a}\right) \; ,
\end{equation}
with $\phi$ being a Gaussian probability density. 
To simplify the notations, we use indices 
from the beginning of the Latin alphabet for measurements 
and the Greek letters for derivatives 
$\frac{\partial}{\partial\mathbf{p}^\mu}$, 
shortened to $\partial_\mu$.
The log-likelihood can then be Taylor expanded 
to the second order to find parameter uncertainties 
by the Cramer-Rao bound 
\citep{Amari2016InformationApplications} 
using the Hessian of the log-likelihood,
\begin{equation}
g_{\mu\nu}(\mathbf{p})= \sum\limits_a \frac{ \partial_\mu f^a\partial_\nu f^a}{\left(\sigma^a\right)^2} \; .
\end{equation}
The above quantity is referred to as the 
Fisher information matrix (FIM).

\subsection{Information geometry}

The simple picture described above can be reinterpreted 
by using information geometry. 
The function $l(\mathbf{p})$ connects 
$\mathcal{M}$ and $\mathcal{N}$, now interpreted 
as manifolds. 
Furthermore, the differential form, i.e., 
$dl = \partial_\mu l d\mathbf{p}^\mu$, 
forms a basis for the cotangent bundle on $\mathcal{N}$, 
labeled as $T^*\mathcal{N}$, 
while the FIM serves as a metric on $\mathcal{N}$.
Here, we note that the appearance of the same 
index ($\mu$) more than once in the 
mathematical expression indicates summation 
with respect to that index, and 
we follow this convention from now on. 
The functional form of the log-likelihood is then 
used to induce a metric on the parameter space 
$\mathcal{M}$. 
This is achieved by computing the expectation 
value taken with respect to $\mathcal{N}$: 
$g \equiv E[dl \otimes dl]$ 
\citep{Amari2016InformationApplications}. 
The pullback operation $l^*$ then induces 
a metric $g(\mathbf{p})\in (T^*\mathcal{M})^2$ 
on $\mathcal{M}$, 
as $g(\mathbf{p})=g_{\mu\nu} d\mathbf{p}^\mu \otimes d\mathbf{p}^\nu=E[\partial_\mu l \partial_\nu l] d\mathbf{p}^\mu \otimes d\mathbf{p}^\nu =l^* g$. 
This procedure equips the model manifold 
$\mathcal{M}$ with a tangent bundle spanned 
by the basis $\partial_\mu\in T\mathcal{M}$ 
and its cotangent bundle spanned by the 
dual basis $d\mathbf{p}^\mu\in T^*\mathcal{M}$. 
Since the normal family is a subset of the 
exponential family, the model manifold 
$\mathcal{M}$ is therefore a submanifold 
embedded in $\mathcal{N}$ and belongs to 
the curved exponential family \citep{Amari1982}.

In differential geometry, tangent spaces 
of nearby points in $\mathcal{M}$ are connected 
via the covariant derivative, 
usually expressed as $\nabla_X$ 
with an arbitrary direction $X$. 
The action of the covariant derivative 
on a tangent vector 
$Y\in T\mathcal{M}$ is simply given by 
$\nabla_X(Y)=\nabla_X(Y^\mu\partial_\mu)=X^\nu\partial_\nu(Y^\mu)\partial_\mu+\Gamma^\kappa_{\mu\nu}X^\mu Y^\nu\partial_\kappa$. 
The quantity $\Gamma^\kappa_{\mu\nu}$ stands for 
the Christoffel symbol when the metric-compatible 
connection with the condition $\nabla_X(g)=0$  
is chosen 
(for details see, e.g., Ref.~\citep{Lee2018IntroductionManifolds}). 
For the FIM, the Christoffel symbols are given by
\begin{equation}
\Gamma^\kappa_{\mu\nu}(\mathbf{p}) = 
g^{\kappa\rho}\sum\limits_a \frac{\partial_\rho f^a\partial_{\mu\nu} f^a}{\left(\sigma^a\right)^2} \; ,
\label{eq:Christoffel}
\end{equation}
where {$g^{\kappa\rho} = (g^{-1})_{\kappa\rho}$} 
denotes the inverse of the metric.

Additionally, along the geodesic, we compute 
the Riemann curvature tensor and the scalar curvature. 
We implement the Riemann curvature tensor, 
defined for $X,Y,Z\in T\mathcal{M}$, 
as $R(X,Y)Z=[\nabla_X,\nabla_Y]Z-\nabla_{[X,Y]}Z$. 
The components of the Riemann tensor 
are expressed as
\begin{equation}
    R_{\mu\nu\rho\kappa}=
   \sum\limits_{ab} P^{ab}\left(\partial_{\mu\rho}\frac{f^a}{\sigma^a}\partial_{\nu\kappa}\frac{f^b}{\sigma^b}-\partial_{\mu\kappa}\frac{f^a}{\sigma^a}\partial_{\nu\rho}\frac{f^b}{\sigma^b}\right) 
\; ,
\end{equation}
where $P^{ab}$ denotes the projection operator
\begin{equation}
P^{ab}=\delta^{ab}-g^{\mu\nu}\partial_\mu \frac{f^a}{\sigma^a} \partial_\nu \frac{f^b}{\sigma^b}
\; .
\end{equation}
The Ricci scalar (or scalar curvature) 
is computed simply as 
\begin{equation}
R_{\mu\nu\rho\kappa} g^{\mu\rho} g^{\nu\kappa}
\; . 
\end{equation}

\subsection{The manifold boundary approximation method}

Complex models might have large parameter 
uncertainties, i.e., a large covariance matrix. 
In the cases where the covariance matrix, 
and therefore the corresponding FIM, has a spectrum 
spanning many orders of magnitude 
\citep{Machta2013ParameterModels}, model reduction 
procedures can improve parameter estimates.
The MBAM \citep{Transtrum.2014} allows 
for better constraining parameters of such models 
across many physical disciplines \citep{Transtrum.2015}. 
The method computes the geodesic by solving 
the geodesic equation 
$\nabla_{\mathbf{\dot p}} \mathbf{\dot p} = 0$ 
by starting from the best-fitting (bf) point 
in the model manifold, 
$\mathbf{p}_\text{bf} \equiv \mathbf{p}_\text{bf}^\mu\partial_\mu$. 
Note that the dot on $\mathbf{p}$ represents 
the differentiation with respect to the 
affine parametrization of the geodesic. 
The geodesic equation, written in parameter 
components as 
\begin{equation}
    \mathbf{\ddot p}^\kappa+\Gamma^\kappa_{\mu\nu}\mathbf{\dot p}^\mu\mathbf{\dot p}^\nu=0,\label{eq:geodesic_comp}
\end{equation}
is solved with the $\mathbf{\dot p}$ 
initial conditions pointing in the direction 
of the FIM eigenvector $v^0$, corresponding 
to its smallest eigenvalue. 
This is the largest eigenvalue of the covariance 
matrix and the biggest contributor to the uncertainty 
of the derived model parameters.
The behavior of $v^0$ is followed along the 
geodesic until the parameter, or combination of 
parameters, contributing most to $v^0$ can be easily 
determined. 
This parameter is then eliminated from the model, 
resulting in a simpler model with smaller parameter 
uncertainties. This procedure can be repeated 
as long as the reduced model describes 
the data set sufficiently well.

\section{Illustrative calculation\label{sec:numerics}}

The density-dependent point-coupling (DD-PC1) 
interaction \citep{Niksic2008Finite-Interactions} 
is a semi-empirical relativistic EDF 
that involves the point coupling \citep{Fin.04} 
and has been used in many contemporary 
studies of nuclear structure and dynamics. 
The DD-PC1 functional explicitly includes nucleon degrees 
of freedom and considers only second-order interaction 
terms. Its applicability to a wide range 
of atomic nuclei has been demonstrated, e.g., in 
Refs.~\citep{Agbemava2014GlobalUncertainties,Agbemava2015CovariantNuclei}.

We use the Dirac mass and energy density data 
shown in Table~\ref{tab:table1} to constrain 
the density-dependent coupling constants 
of the DD-PC1 functional, 
$\alpha_s(\rho)$, $\alpha_v(\rho)$, 
and $\alpha_{tv}(\rho)$, 
modeled as \citep{Niksic.2016, Niksic.2017}
\begin{equation}
\alpha_i = a_i + 
\left(b_i+c_i\frac{\rho}{\rho_\text{sat}}\right)
e^{-d_i \frac{\rho}{\rho_\text{sat}} }
\; ,
\quad 
i\in\{s,v,tv\}
\; ,
\end{equation}
where the indices 
$i=s$, $v$, and $tv$ correspond to the 
isoscalar-scalar, 
isoscalar-vector, and isovector-vector channels, 
respectively, 
while $\rho_\text{sat}$ is the saturation density. 
In this paper, we take a closer look at the reduced 
version of the model with 
$\alpha_{tv}=0$ and $c_v=0$, which results in 
a seven-parameter model involving  
$a_s$, $b_s$, $c_s$, $d_s$, $a_v$, $b_v$, and $d_v$. 

\begin{table}[h]
\caption{\label{tab:table1}%
Pseudodata for infinite symmetric nuclear matter 
used to compute the best-fitting solution for 
the energy density functional. The adopted error 
for the $y$ points is $10\%$ for energy and 
$2\%$ for the Dirac mass.
}
\begin{ruledtabular}
\begin{tabular}{cccc}
Index & $\rho_v$ fm$^{-3}$ &
\textrm{$y$ \footnote{first row is in $M_D/m$, otherwise in MeV units.}} &
\textrm{$\sigma_y$}\\
\colrule
1& 0.152 &  0.58 &  0.055\\  
2& 0.04  & $-$6.48 &  0.648\\ 
3& 0.08  & $-$12.13&  1.213\\ 
4& 0.12  & $-$15.04&  1.504\\
5& 0.16  & $-$16.  &  1.6  \\
6& 0.2   & $-$15.09& 1.509 \\ 
7& 0.24  & $-$12.88& 1.288 \\  
8& 0.32  & $-$5.03 &0.503
\end{tabular}
\end{ruledtabular}
\end{table}

\subsection{Numerical implementation}

We solve the equation for the Dirac mass $M_D$, 
which is given by \cite{Niksic.2016}
\begin{equation}
\label{eq:MD}
M_D = m + \alpha_s \rho_s \; ,
\end{equation}
where $m$ is the bare nucleon mass, 
and the scalar density $\rho_s$ 
\begin{equation}
\rho_s = \frac{2}{\pi^2}M_D\int_0^{p_F}{\frac{x^2dx}{\sqrt{x^2+M_D^2}}}
\; ,
\end{equation}
with $p_F$ being the Fermi momentum
\begin{equation}
p_F(\rho_v)=\left(\frac{3}{2}\rho_v\pi^2\right)^\frac{1}{3}
\; .
\end{equation}
Equation (\ref{eq:MD}) is solved numerically 
by using the Newton-Raphson algorithm. 
We also tested Halley's method, 
but found no improvement of the results 
in accuracy.

Upon finding $M_D$, we compute the 
binding energy of symmetric nuclear matter: 
\begin{equation}
E_a = 
\frac{2}{\pi^2} \int_0^{p_F}{\frac{x^4dx}{\sqrt{x^2+M_D^2}}}
+ m(\rho_s-\rho_v)+\frac{1}{2} 
\alpha_s\rho_s^2+\frac{1}{2}\alpha_v\rho_v^2 
\; .
\end{equation}
The best-fitting DD-PC1 parameter 
set is then found by computing the least-square 
solution to the set of measurements of $M_D/m$ 
and $E_a$ presented in Table~\ref{tab:table1} 
(see Ref.~\citep{Niksic.2017}). 
Differential equations are solved with the 
aid of the SCIPY implementation of the ordinary 
differential equation integration ({odeint}) 
library \citep{2020SciPy-NMeth}.
These values are then used to compute the FIM 
and the Christoffel symbols using algorithmic
differentiation implemented via the AUTOGRAD
package. We thus eliminate numerical errors due to the
approximations arising from numerical differentiations.

\begin{figure*}[ht]
\begin{center}
 \includegraphics[width=.9\linewidth]
{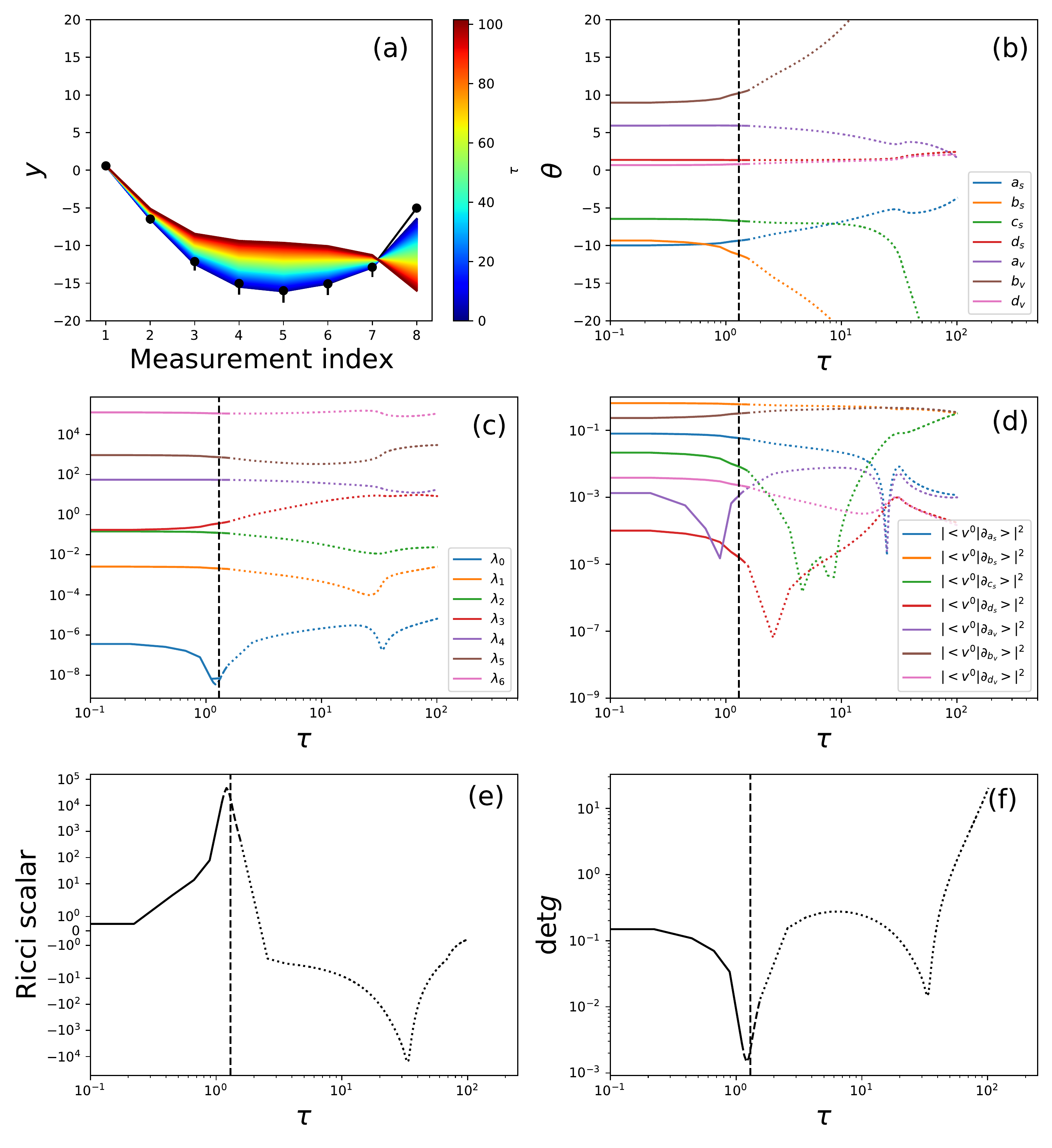}
\caption{Results of extrapolating 
the geodesic after the $\det g=0$ point. 
Shown are (a) the behavior of the evaluated model 
for different $\tau$'s along the geodesic, 
(b) the model parameters, 
(c) the FIM eigenvalues as functions of $\tau$, 
(d) the squares of the FIM eigenvector $v^0$ components, 
(e) the Ricci scalar, and 
(f) the FIM determinant along the geodesic. 
Solid, dashed, and dotted lines stand for, 
respectively, 
the initial odeint solutions, 
the linear interpolation, and 
the values derived using odeint starting from 
the endpoint of the interpolated solutions.}
\label{fig:interp}
\end{center}
\end{figure*}

\begin{figure*}[ht]
\begin{center}
\includegraphics[width=\linewidth]
{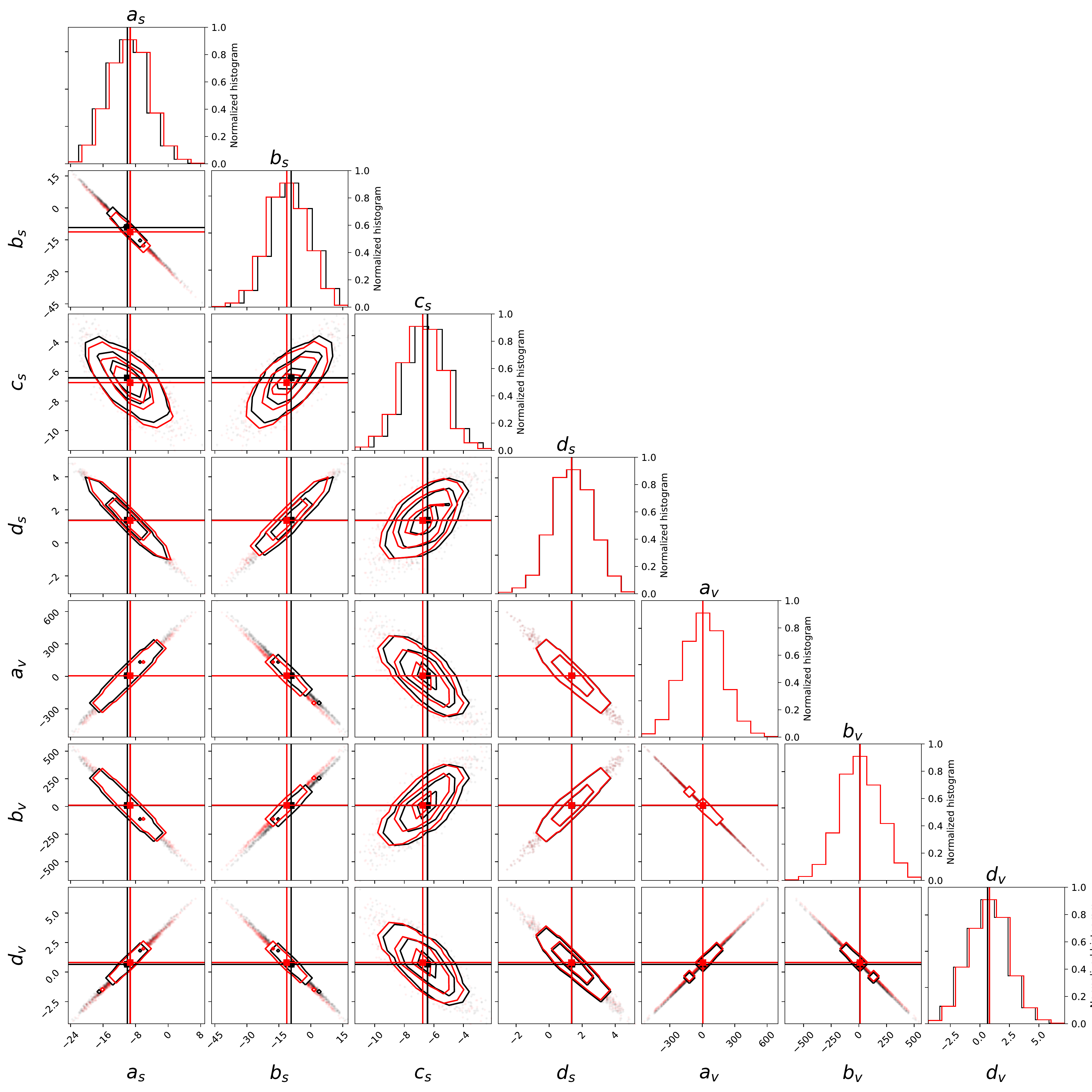}
\caption{\label{fig:MBAMprop_pars}
Monte Carlo simulated sample parameters 
using the best-fitting covariance matrix 
(black symbols and contours) and 
its propagation towards $\tau=1.3$ 
along the geodesic using the 
Jacobi equation (\ref{eq:jacobi}) 
(red symbols and contours).}
\end{center}
\end{figure*}

\section{Investigating stability of the MBAM method\label{sec:stability}}

In some cases, the numerical integration of the 
geodesic equation might slow down, or even fail. 
This behavior is due to the divergence of the metric 
tensor determinant that implicitly appears 
in the geodesic equation (\ref{eq:geodesic_comp}) 
through the metric inverse necessary for 
computing the Christoffel symbols 
[see Eq.~(\ref{eq:Christoffel})]. 
However, this divergent behavior is confined to only 
a small region in the parameter space and therefore 
it might be easily missed by choosing too imprecise 
an integrator. 
Therefore, in Sec.~\ref{sec:extrap}, 
we investigate the impact of the size of the 
integration step on the MBAM procedure by 
artificially extrapolating the geodesic beyond the 
divergent region in the parameter space.
Moreover, as the parameter uncertainties 
become larger, small perturbations to the starting 
point of the geodesic might influence 
the end result of the MBAM. 
{In Sec.~\ref{sec:uncertainty_prop}, 
we describe the impact of parameter uncertainties 
on the MBAM conclusions for the nuclear 
EDF DD-PC1 by numerical error propagation 
of the MBAM geodesics.}
{Finally, in Sec.~\ref{sec:reparametrization} 
we investigate the impact of using a common, 
physically motivated restrictive reparametrization 
of the DD-PC coupling constants on the MBAM 
model manifold.}

\subsection{Geodesic extrapolation\label{sec:extrap}}

We extrapolate the geodesic by using the 
last point having $\det g>0$ 
(labeled as $\tau_2$) and the point before 
it ($\tau_1$). 
We first extrapolate $\tau(t)=\tau_1 (1-t) +\tau_2 t$ 
for $t>0$, i.e., a straight line joining 
$\tau_1$ and $\tau_2$. 
We then compute the $\mathbf{p}^\mu(t)$ and 
$\mathbf{\dot p}^\mu(t)$ using their corresponding 
values at $\tau_1$ and $\tau_2$ as 
\begin{align}
& \mathbf{p}^\mu (t) = \mathbf{p}^\mu(\tau_1) (1-t)+\mathbf{p}^\mu(\tau_2) t \\
& \mathbf{\dot p}^\mu (t) = \mathbf{\dot p}^\mu(\tau_1) (1-t)+\mathbf{\dot p}^\mu(\tau_2) t \; .
\end{align}
This procedure produces a linear extrapolation 
of the geodesics in the region where the geodesic 
equation does not hold because $\det g = 0$. 
The variable $t$ is just an interpolation parameter, 
not connected to $\tau$, so $\mathbf{\dot p}$ 
is not coupled as $d\mathbf{p}/d\tau$ in this region. 
We find that one can safely continue 
integrating the geodesic equation after $t=2$, 
where there are no more singularities 
along the path.

In Fig.~\ref{fig:interp}, 
the resulting model parameters along the 
extended geodesic (a), 
the corresponding model evaluation (b), 
the FIM eigenvalues (c), the $v^0$ eigenvector (d), 
the Ricci scalar (e), and the metric determinant (f)
are shown. 
After the $t=2$ point along the extrapolated 
geodesic, the metric tensor determinant starts 
to rise again. 
In the same figure, 
the linearly extrapolated 
geodesic, corresponding to the small 
region $\tau\in [\tau(t=1), \tau(t=2)]$, 
is shown with dashed lines. 
The extrapolated geodesic computed using MBAM 
continuation starting from the point $\tau(t=2)$ 
is shown with dotted lines.
The initial {odeint} 
solutions (solid lines), which produce 
results for a few points 
after $\tau=1.3$, differ significantly 
from the interpolated solution, 
indicating numerical problems due to singularity. 
Upon restarting the {odeint} procedure after 
the singular region, we find that the MBAM 
solution yields different contributions 
to the $v^0$ eigenvector, 
indicating an equal contribution of 
$\partial_{b_s}$, $\partial_{c_s}$, 
and $\partial_{b_v}$ directions, 
while before $\tau=1.3$, the MBAM method finds 
that the most significant contribution is from $\partial_{b_s}$. 
The Ricci scalar diverges at $\tau\sim 1.3$, 
but starts to fall and change signs at $\tau>1.3$. 
Since the Ricci scalar is related to the 
volume element, its divergence to positive 
values would produce a compressed region 
of the parameter manifold, which begins 
to expand after the singularity.

The conclusion drawn from 
the results given in Fig.~\ref{fig:interp} 
is that one must be careful 
with the models where the metric tensor 
determinant shows significant variations, 
as choosing too big steps for the odeint 
integrator might result in ``skipping'' 
to another portion of the parameter space 
and continuing along it. This yields 
completely different contributions 
to the FIM eigenvector corresponding 
to its smallest eigenvalue and hence 
might lead to a completely different model 
reduction than expected from the simple MBAM case.

\begin{figure}[ht]
\begin{center}
 \includegraphics[width=\linewidth]
{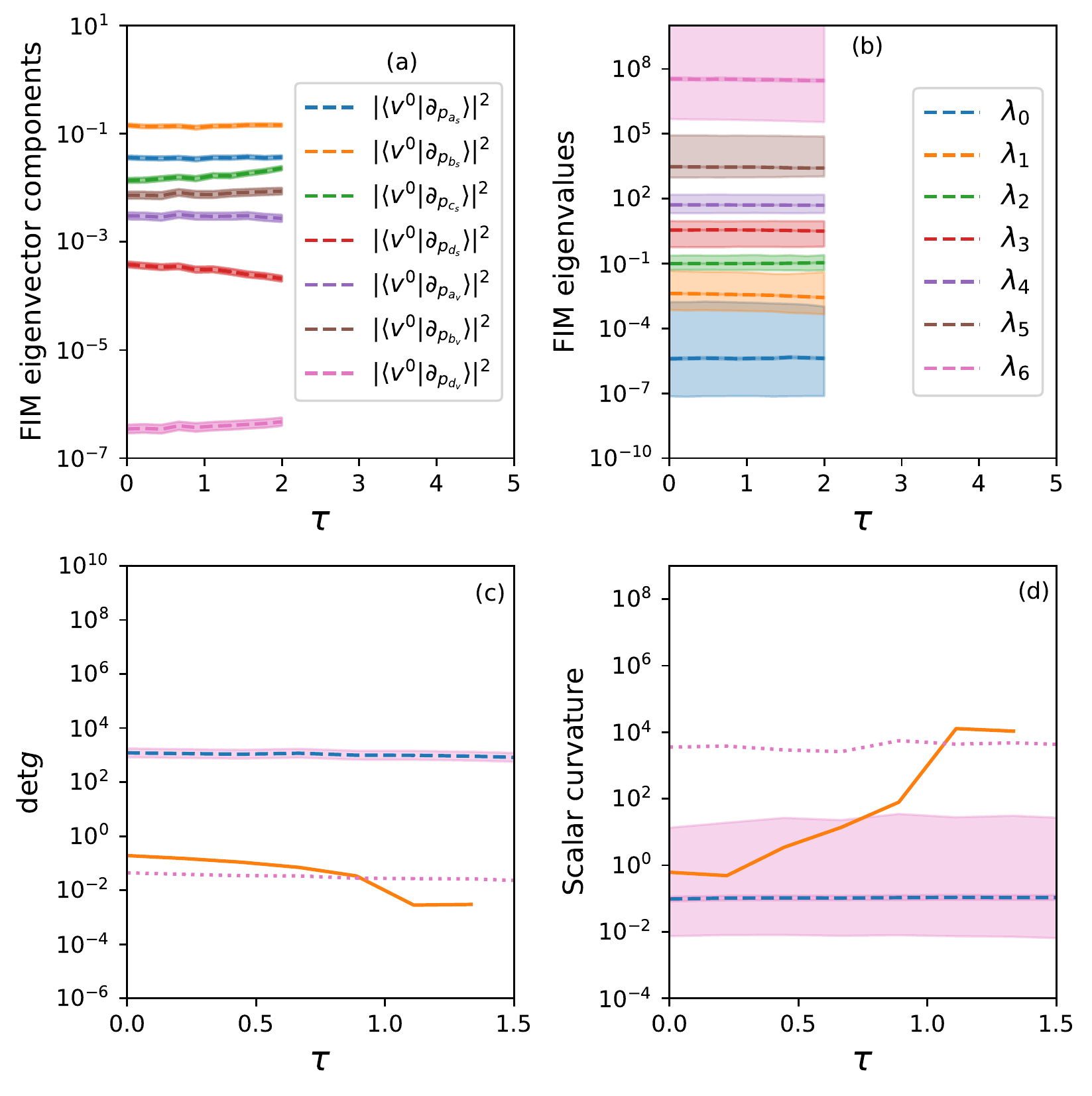}
\caption{Monte Carlo simulations of 
uncertainty propagation 
using the Jacobi equation (\ref{eq:jacobi}). 
Shown are the median and its uncertainty 
derived using 1300 simulated samples starting 
from the best-fitting point. 
The figure shows 
(a) the simulated FIM 
$v^0$ eigenvector components squared, 
(b) FIM eigenvalues, 
(c) FIM determinant, and 
(d) scalar curvature. 
The shaded areas correspond to the $1\sigma$ 
percentile interval, while the dotted lines 
in panels (c) and (d) additionally show the 
5th and the 95th percentiles, respectively. 
Solid orange lines in panels (c) and (d) 
stand for the respective quantities 
computed along the path of the MBAM geodesic. }
\label{fig:MBAMerrors}
\end{center}
\end{figure}

\begin{figure*}[ht]
\begin{center}
\includegraphics[width=\linewidth]
{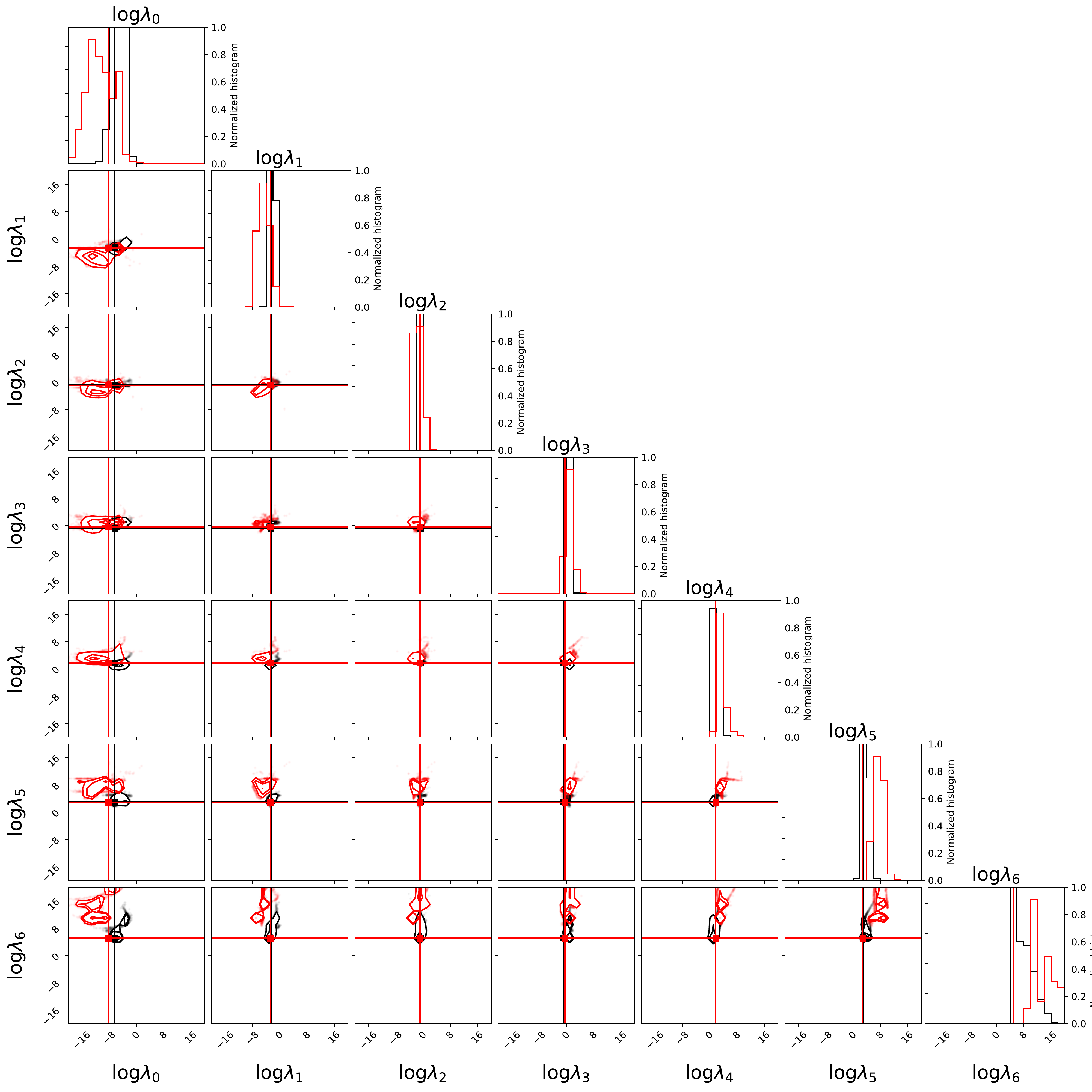}
\caption{
Same as Fig.~\ref{fig:MBAMprop_pars}, 
but for Monte Carlo simulated sample (base 10) 
logarithms of the eigenvalues of the FIM.}
\label{fig:mfig:MBAMprop_eigvals}
\end{center}
\end{figure*}

\begin{figure*}[ht]
\begin{center}
\includegraphics[width=\linewidth]
{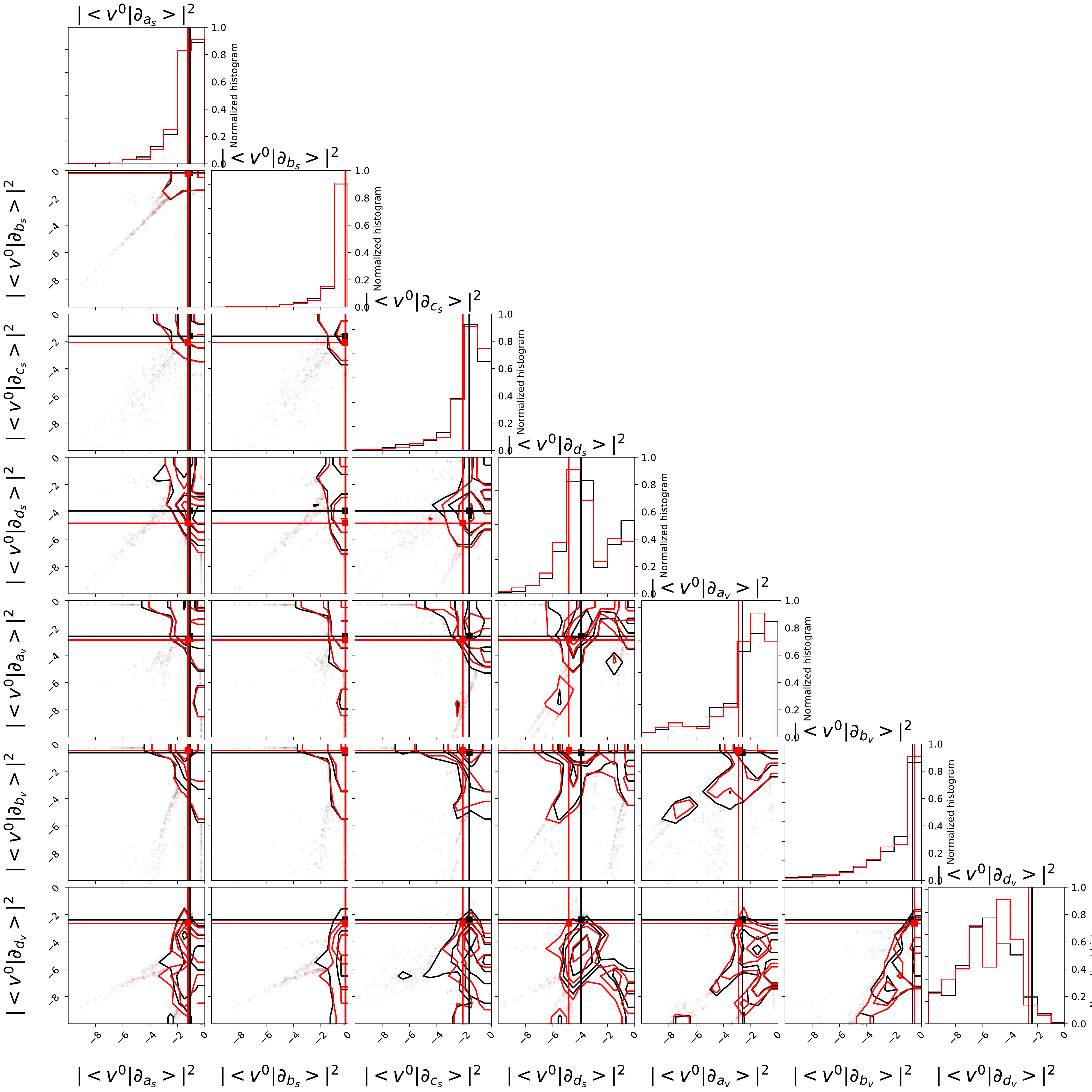}
\caption{
Same as Fig.~\ref{fig:MBAMprop_pars}, 
but for Monte Carlo simulated sample components 
of the FIM $v^0$ eigenvector.}
\label{fig:fig:MBAMprop_v0}
\end{center}
\end{figure*}

\subsection{Parameter uncertainties\label{sec:uncertainty_prop}}

Further extension of the basic model 
might be the propagation of its parameter 
uncertainties, and this can be facilitated 
by looking into how the uncertainties 
of the best-fitting parameters propagate 
along the geodesics. 
For this purpose, we perform Monte Carlo simulations. 
To analyze the error propagation one would 
have to compute the geodesic equation many times, 
which is not cost-efficient. 
We, therefore, adopt a simplified approach 
that makes use of the Jacobi equation, which 
computes differences $\delta\mathbf{p}$ 
between neighboring geodesics 
along the already computed MBAM geodesic. 

We use the covariance matrix $\Sigma$ to produce 
Monte Carlo simulations of $\delta\mathbf{p}$ 
from the normal distribution, 
$\delta\mathbf{p}\sim \mathcal{N}(0,\Sigma)$. 
For each simulated $\delta\mathbf{p}$, 
we compute its propagation by 
using the Jacobi equation
\begin{equation}
\label{eq:jacobi}
\delta\mathbf{\ddot p}^\mu+R^\mu_{\alpha\nu\beta} \mathbf{\dot p}^\alpha \mathbf{\dot p}^\beta \delta\mathbf{p}^\nu=0 \; .
\end{equation}
We find 1300 points to sample the 
DD-PC1 parameter space reasonably well. 
Figure~\ref{fig:MBAMprop_pars} shows 
the distributions of the parameters 
at the beginning (denoted by black symbols 
and contours) and at $\tau=1.3$ 
(red symbols and contours). 
These two distributions are almost identical 
since the simulated parameters are more 
dispersed than the gradual changes 
in parameter values along the geodesic.

Even though the parameter uncertainties 
in the full model are large, we can estimate 
the error on the eigensolutions of 
the FIM along the geodesic. 
We do this by computing the FIM 
for every simulated point propagated along 
the best-fitting geodesic to various values 
of $\tau$ using the Jacobi equation. 
The results of this procedure are shown 
in Fig.~\ref{fig:MBAMerrors}. 
The top panels show the median and the 
corresponding $1\sigma$ confidence interval 
of the eigensolutions, computed using the 
16th and the 84th percentile. The simulated FIM 
$v^0$ eigenvector components 
squared are shown in Fig.~\ref{fig:MBAMerrors}(a) 
and the FIM eigenvalues are shown 
in Fig.~\ref{fig:MBAMerrors}(b) for each $\tau$.
We see that, while the results using 
the simulated sample are consistently ordered 
when compared to the MBAM solution, 
there is a small offset between the median 
and the MBAM solution. 
Figures~\ref{fig:MBAMerrors}(c) and 
\ref{fig:MBAMerrors}(d) show the median and 
the $1\sigma$ confidence interval for the FIM 
determinant and the scalar curvature, respectively. 
The simulated scalar curvature and the metric 
determinant along the geodesic show a larger 
variation in their values along the geodesic. 
In these panels we additionally show the 
FIM determinant and the scalar curvature along 
the best-fitting geodesic by the solid orange lines. 
There is a large discrepancy between the behavior 
of the median of the simulated quantities 
and the behavior of the quantities along the 
best-fitting geodesic. 
In Fig.~\ref{fig:MBAMerrors}(c) 
[Fig.~\ref{fig:MBAMerrors}(d)], 
we see that these quantities along the best-fitting 
geodesic are comparable to the 5th (95th) percentile 
of $\det g$ (scalar curvature), shown as dotted lines. 
This behavior indicates that only the geodesics 
starting at the vicinity 
of the best-fitting point encounter 
the region corresponding to $\det g=0$.

Furthermore, 
in Figs.~\ref{fig:mfig:MBAMprop_eigvals} 
and \ref{fig:fig:MBAMprop_v0} we show, 
respectively, the 
distributions of eigenvalues and components 
of $v^0$ at the beginning and at the end of 
the geodesic. 
These large differences in eigenvalues 
and eigenvector components propagating along 
the geodesic are in stark contrast to the 
parameter values in Fig. \ref{fig:MBAMprop_pars}. 
The discrepancies presented 
in Figs. \ref{fig:MBAMerrors}, 
\ref{fig:mfig:MBAMprop_eigvals}, 
and \ref{fig:fig:MBAMprop_v0} can be explained 
by the sensitivity of the FIM eigenproblem 
to small changes in DD-PC1 parameters, 
since diagonalization results are 
not expected to change linearly with inputs. 
We conclude that the offset is due to the 
non-Gaussianity of the distribution of eigenvalues 
and $v^0$ components, which arises even though 
the parameters were sampled using the normal 
distribution. Even though there is a change in 
the shape of these distributions, the overall 
qualitative MBAM conclusions remain the 
same along the geodesic.

\begin{figure}[h]
\begin{center}
 \includegraphics[width=\linewidth]
{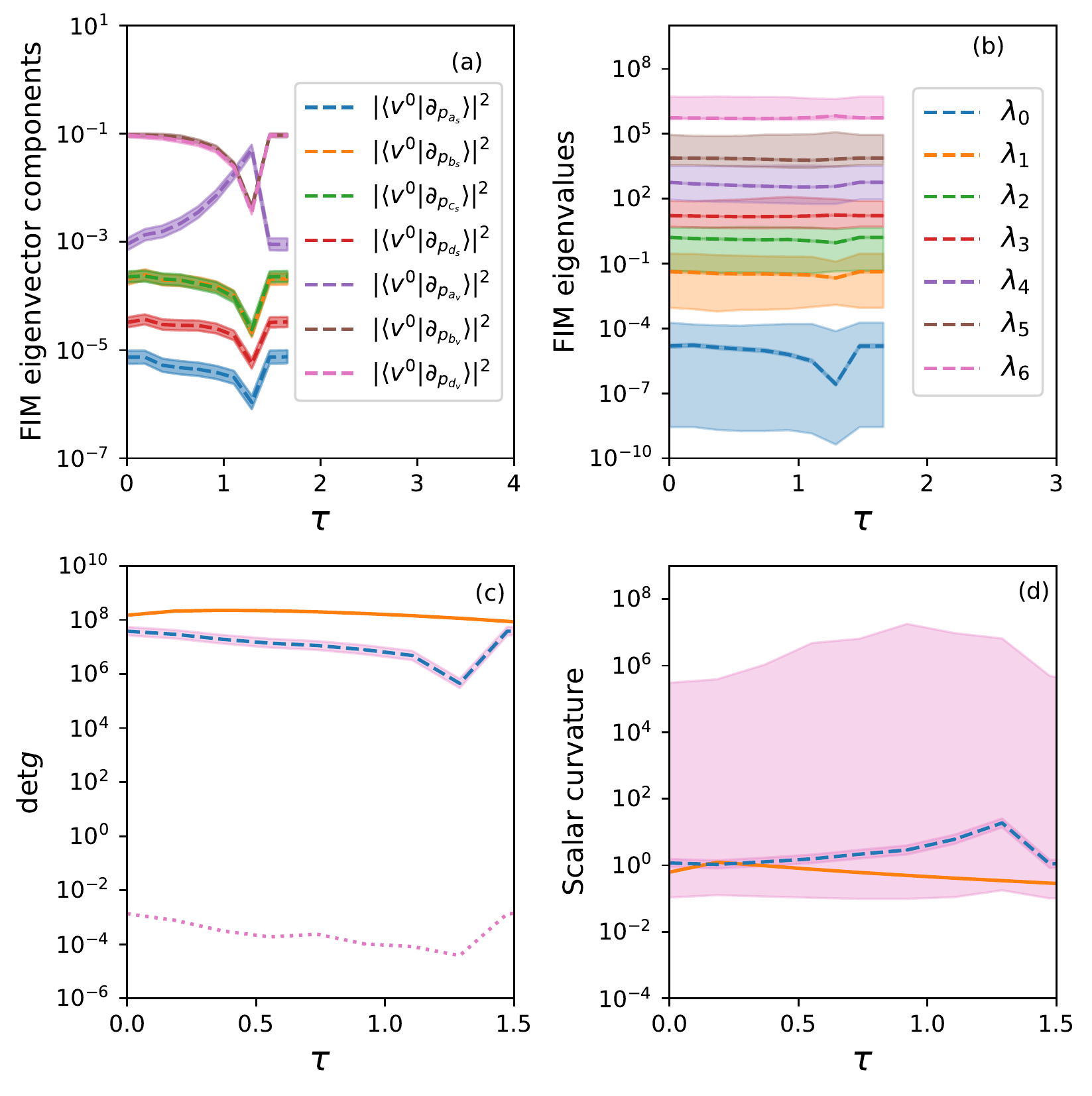}
\caption{Same as Fig.~\ref{fig:MBAMerrors}, 
but for the reparametrized model described 
in Sec.~\ref{sec:reparametrization}.}
\label{fig:MBAMerrorsreparam}
\end{center}
\end{figure}

\subsection{{Model reparametrization}\label{sec:reparametrization}}

The authors of Ref.~\citep{Niksic.2016} 
have considered an exponential reparametrization 
of the seven-parameter DD-PC1 coupling 
constants centered at their best-fitting values 
\cite{Niksic2008Finite-Interactions}. 
This reparametrization transformation can 
be schematically represented as a vector
\begin{align}
\mathbf{p}(\mathbf{\tilde p})=
\begin{pmatrix} 
    a_s(p_{a_s}) \\
    b_s(p_{b_s}) \\
    c_s(p_{c_s}) \\
    d_s(p_{d_s}) \\
    a_v(p_{a_v}) \\
    b_v(p_{b_v}) \\
    d_v(p_{d_v}) 
\end{pmatrix}
=
\begin{pmatrix}
     a_{s,\text{bf}}\,e^{-p_{a_s}}\\
     b_{s,\text{bf}}\,e^{-p_{b_s}}\\
     c_{s,\text{bf}}\,e^{-p_{c_s}}\\
     d_{s,\text{bf}}\,e^{-p_{d_s}}\\
     a_{v,\text{bf}}\,e^{-p_{a_v}}\\
     b_{v,\text{bf}}\,e^{-p_{b_v}}\\
     d_{v,\text{bf}}\,e^{-p_{d_v}}
\end{pmatrix}
\; ,
\label{eq:repar}
\end{align}
where $\mathbf{\tilde p}$ indicates 
the multivariate distribution of parameters 
$p_{a_s},\cdots,p_{d_v}$, 
and the quantities such as 
$a_{s,\text{bf}}$, $b_{s,\text{bf}}$, etc., 
stand for the best-fitting parameter values. 
The exponential form of the coupling constants 
is chosen by the constraints 
(i) that the new parameters in the geodesic equation 
are dimensionless and 
(ii) that the exponential form prevents the 
coupling functions $a_s$ and $a_v$ from changing 
sign along the geodesic path, thus confining them 
in the region described by the inequalities 
$\alpha_s<0$ and $\alpha_v>0$. 
Using these constraints the scalar mean-field 
potential remains attractive and the vector 
mean-field remains repulsive for all allowed 
parameter values \cite{Niksic.2016}.

\begin{figure*}[ht]
\begin{center}
 \includegraphics[width=\linewidth]
{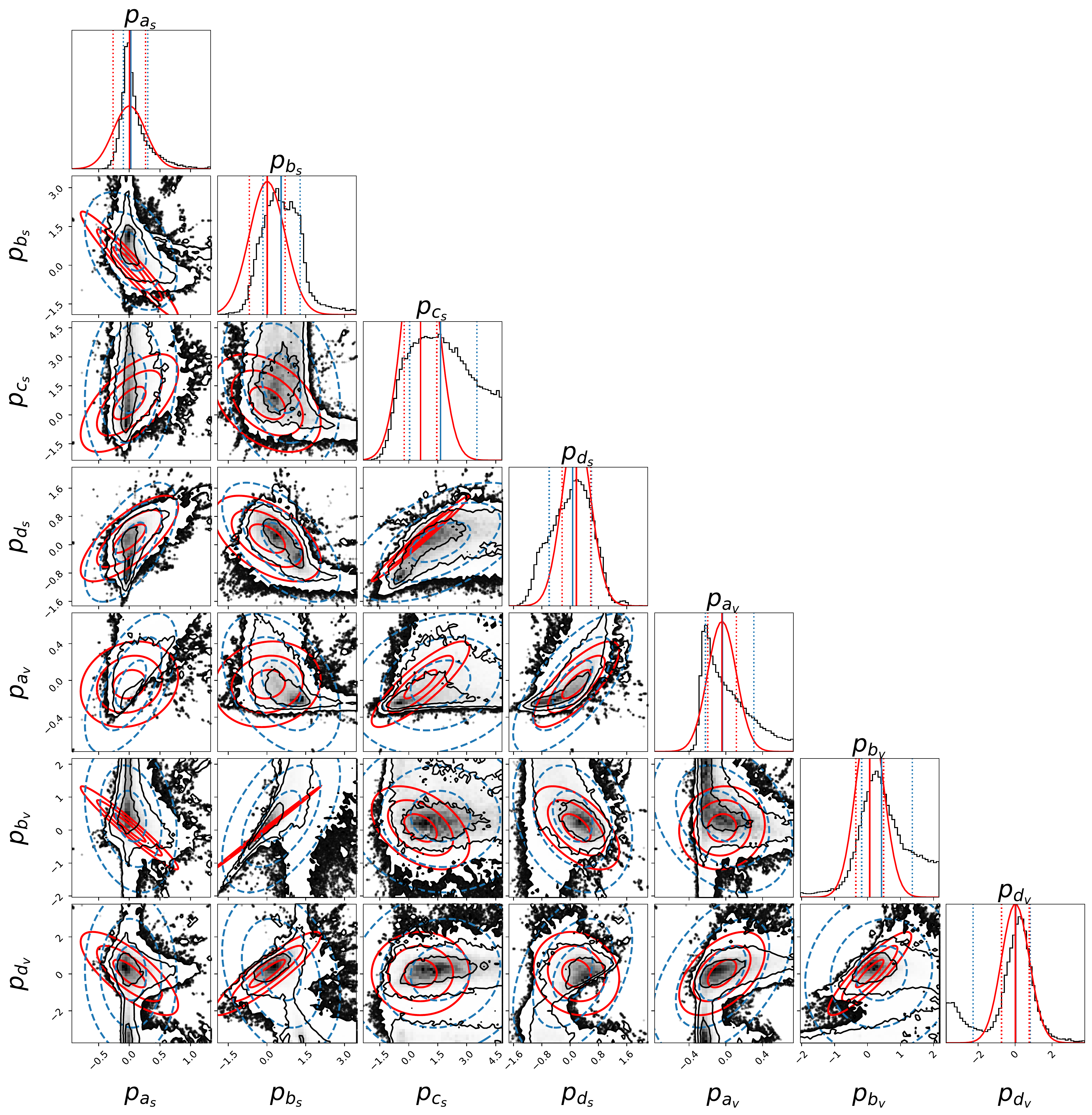}
\caption{Monte Carlo simulations of posterior 
distributions of the error estimates for the 
reparametrized model, based on the MCMC algorithm. 
The figure shows the $1\sigma$, $2\sigma$, 
and $3\sigma$ covariance ellipses in red, 
as estimated from the FIM inverse, and the 
estimates of the covariance ellipses based 
on the MCMC sample points in blue.}
\label{fig:mcmc}
\end{center}
\end{figure*}

We repeat the Monte Carlo analysis described in 
Sec.~\ref{sec:uncertainty_prop} for 
the reparametrized model. 
The resulting error estimates are shown 
in Fig.~\ref{fig:MBAMerrorsreparam} 
in the same manner as in Fig. \ref{fig:MBAMerrors}. 
By comparing the two figures panel-by-panel, 
we conclude that both methods produce MBAM 
geodesics that are stable under perturbations, 
even though the two FIMs do not behave in 
the same way along their respective geodesics. 
The reparametrized FIM determinant and the 
Ricci scalar change gradually, compared 
to the initial model.

One may ask whether this discrepancy is due to 
using a too simplistic description of the 
reparametrized distributions. 
We then employ the Bayesian statistics to 
check whether the multivariate 
distribution $\mathbf{\tilde p}$ 
has pronounced non-Gaussian features. 
To this end, we use the Markov chain 
Monte Carlo (MCMC) technique to sample the 
$\chi^2$ posterior distribution, 
as implemented in the package EMCEE 
\cite{2013PASP..125..306F}. 
In Fig.~\ref{fig:mcmc} we show the behavior 
of the chosen 200 Markov chains as two-dimensional 
sections of the parameter space. 
The chains have been run for a long enough time 
to avoid the initial ``burn-in'' phase characteristic 
of the algorithm during which they follow mostly 
the (uniform) prior distribution instead of sampling 
the $\chi^2$ posterior distribution. 
From the fact that the classical covariance ellipses 
(represented by red contours 
in Fig.~\ref{fig:mcmc}) are well aligned 
with the MCMC estimates, 
we conclude that one can proceed with using the 
simple Monte Carlo Gaussian mock sample for error 
propagation instead of the computationally more 
expensive Bayesian MCMC mock sample.

The theoretical argument for the discrepancy 
between the two geodesics is based 
on the properties of the applied transformation. 
Since the exponential transformations 
are not bijections, the geodesics on the 
manifold spanned by $\mathbf{\tilde p}$ 
need not have the same behavior as the geodesics 
on the manifold spanned by $\mathbf{p}$. 
To better understand the connection between 
these two geodesics, we derive the FIM 
determinant on the $\mathbf{\tilde p}$-manifold 
by using the transformation of Eq.~(\ref{eq:repar}), 
\begin{widetext} 
\begin{equation}
\det g(\mathbf{\tilde p})
= a_{s}^2(p_{a_s})b_{s}^2(p_{b_s})c_{s}^2(p_{c_s})d_{s}^2(p_{d_s})a_{v}^2(p_{a_v})b_{v}^2(p_{b_v})d_{v}^2(p_{d_v})\det g(\mathbf{p} (\mathbf{\tilde p}))
\; .
\label{eq:detgp}
\end{equation}
\end{widetext}
The determinant of the metric is not an invariant 
quantity under reparametrizations, and 
hence the additional 
multiplicative scaling is required. 
Equation (\ref{eq:detgp}) shows that, 
if the value of  $\det g$ approaches zero 
for particular values of $\mathbf{p}$, 
both geodesics terminate. 
However, additional singularities appear 
if any of the coupling constants are allowed 
to change sign along a particular geodesic. 
In contrast to the FIM determinant, 
the Ricci scalar is not affected 
by reparametrizations. 
The scalar curvature distributions 
for different points on the geodesic 
in Fig.~\ref{fig:MBAMerrorsreparam}(d) 
do not have the same values as those 
in Fig.~\ref{fig:MBAMerrors}(d). 
The effects of reparametrizations 
on the scalar curvature 
can be clearly seen from the comparison 
between these figures.

The general conclusion is, therefore, 
that the MBAM method is sensitive to 
the way the reparametrization is made, 
as has been shown above in the case of 
the reparametrization tied to domain restrictions. 
This is related to the fact that 
different reparametrizations do not 
lead to the same, but similar, models 
describing the common physical problem. 
Since the EDF has an arbitrarily chosen 
functional form, there is no {\textit{a priori}} 
way of identifying which parametrization is optimal.
This sensitivity only emphasizes the fact that 
different reparametrizations may describe similar, 
but inherently different, physical models. 
Choosing a particular EDF parametrization 
is equivalent to choosing a particular range 
model parameters can take.

\section{Conclusion\label{sec:conclusion}}

Methods of information geometry have been 
applied to investigate the stability of reducing 
the nuclear structure models. 
We have constrained the error estimates of 
the MBAM solutions by means of the Monte Carlo 
simulations. 
In the illustrative application 
to the DD-PC1 model of the nuclear EDF, 
it has been found that the main conclusions 
obtained by using the MBAM method are stable under 
the variation of the parameters within 
the $1\sigma$ confidence interval of the 
best-fitting model. 
Moreover, we have found that the end of the 
geodesic occurs when the determinant of the FIM 
approaches zero, thus effectively separating 
the parameter space into two disconnected regions.

Further applications of information geometry 
to nuclear EDFs could be analyzing possible phase 
transitions in models of finite nuclei using scalar 
curvature and their impact on nuclear properties. 
The analysis could even be expanded to include 
an extended temperature-dependent model 
or to look for model instabilities. 
It would be worth investigating whether 
information-theoretic optimizations 
could accelerate computer codes to solve 
nuclear many-body problems. 
Such second-order optimization algorithms, 
as the natural-gradient descent, find optimal 
solutions by taking optimization steps in the 
parameter space informed by the behavior of the FIM.

\section*{Acknowledgments}
This work is financed within 
the Tenure Track Pilot Programme of 
the Croatian Science Foundation and 
the \'Ecole Polytechnique F\'ed\'erale de Lausanne 
and by Project No. TTP-2018-07-3554 
Exotic Nuclear Structure and Dynamics, 
with funds of the Croatian-Swiss Research Programme.

\bibliography{references.bib}
\end{document}